# Magnetic field '*flyby*' measurement using simultaneously magnetometer and accelerometer

Martín Monteiro[a], Cecilia Stari[b], Cecilia Cabeza[c], Arturo C. Marti[d],

[a] Universidad ORT Uruguay; monteiro@ort.edu.uy

[b] Universidad de la República, Uruguay, cstari@fing.edu.uy

[c] Universidad de la República, Uruguay, cecilia@fisica.edu.uy

[d] Universidad de la República, Uruguay, marti@fisica.edu.uy

The spatial dependence of magnetic fields in simple configurations is an usual topic in introductory electromagnetism lessons, both in high school and in university courses. In typical experiments, magnetic fields are obtained taking point-by-point values using a Hall sensor and distances are measured using a ruler. Here, we show how to take advantage of the smartphone capabilities to get simultaneous measures with the built-in accelerometer and magnetometer and to obtain the spatial dependence of magnetic fields. We consider a simple set up consisting of a smartphone mounted on a track whose direction coincides with the axis of a coil. While the smartphone is smoothly accelerated, both the magnetic field and the distance from the center of the coil (integrated numerically from the acceleration values) are simultaneously obtained. This methodology can be easily extended to more complicated setups.

**Simultaneous use of several smartphone sensors**

Recently, the increasing availability and capabilities, and the decreasing cost have contributed to the expansion of *smartphone physics*. Indeed, smartphone sensors, as accelerometer, gyroscope, magnetometer among others, have been successfully employed in diverse physics experiments ranging from mechanics to modern physics (see, for example, this column in past issues of this journal). One relevant aspect that has received little attention is the fact that smartphones allow to obtain simultaneous measures from several sensors. In previous works, the simultaneous use of the accelerometer and the gyroscope has been proposed [1-5]. More recently, the luminosity sensor has been employed together with the orientation sensor to experiment with polarized light [6]. In this work, a simple experience which combines the use of the smartphone magnetometer and the accelerometer is proposed. The magnetic field generated by a current in a coil is measured with a smartphone located over a cart on a rail whose orientation coincides with the axis of the coil. While the smartphone is moving on the track, its position is readily obtained integrating twice the acceleration values obtained from the accelerometer. In this way, with a simple data processing, the magnetic field as a function of the position is obtained and as a by-product also the permeability. These results can be compared with the predictions of the Biot-Savart law.

**Experiment**

Until now a few smartphone-based experiments focusing on electromagnetism have been proposed in the literature (see for example [7-9]). In general, in these experiments the value of the magnetic field is obtained point by point and the distance is measured using a ruler like in traditional approaches. Here, we focus on the axial component of the magnetic field $B$ along the axis of a thin coil carrying a current $I$. According to the Biot Savart law:

$$B = \frac{\mu_0 N I R^2}{2(R^2+y^2)^{3/2}}$$

where $R$ is the mean radius of the coil, $\mu_0$ is the permeability of free space, $N$ is the number of turns, and $y$ is the distance to the center of the coil.

The experimental setup, shown in Fig. 1, consists on a smartphone mounted on a track placed on the axis of a coil aligned with the rail. The coil is made by $N=200$ turns and its mean radius is $R=10.3(2)$ cm. The smartphone is mounted on the track using an aluminum support to avoid magnetic interference. A DC power supply is used to create and electric current in the coil. The current intensity must be chosen such as the maximum magnetic field, that occurs at the center of the coil, do not saturate the sensor. In the present experiment $I=1.83(1)$ A.

The smartphone, a Nexus 5, contains three-axis built-in accelerometer and magnetometer. First of all, we obtain the exact position of the magnetic sensor within smartphone [9]. The position of the smartphone is chosen such that the $y$-axis coincides with the track and that the magnetic sensor is located on the axis of the coil. Secondly, it is necessary to consider that the sensor measures the magnetic field generated by the coil and also additional contributions by the terrestrial magnetic field and other nearby magnetic objects. To obtain the magnetic field produced only by the coil, the background contribution, obtained with zero applied current, must be subtracted. In this experiment the background magnetic field was $72.9 \mu T$. The uncertainties in the values of the magnetic field and the acceleration obtained with the sensors were determined considering standard deviation in a measurement ($0.6 \mu T$ and 0.02 m/s² respectively). The sampling interval was chosen as the maximum rate of the accelerometer ($\Delta t = 0.005 s$).

The smartphone, initially at rest, is pulled and starts to move in a gently accelerated movement. The *App* Androsensor is used to record the $y$-components of both the acceleration and the magnetic field. The measurements obtained using the app, can be processed with a spreadsheet or using a more specific program of data analysis.

The position of the smartphone along the track at the times at which the sensors measured a specific magnetic field value can be obtained integrating the acceleration values. In this case, the Euler method might be easily implemented in an spreadsheet. Given the initial condition at time $t_0$, $v_0 = a_0 = 0$, it is possible to iterate obtaining the variables at time $t_{i+1}$ as a function of the variables at time $t_i$ according to

$$v_{i+1} = v_i + a_i \Delta t$$

$$x_{i+1} = x_i + v_i \Delta t \ .$$

**Results and analysis**

In Fig. 2, the magnetic field as a function of the distance calculated through the numerical integration of the acceleration is displayed. The symbols correspond to the experimental measures while the theoretical model provided by the Biot-Savart law, is presented by the red line. The experimental value of the permeability is obtained from the slope, *s*, of the linear plot shown in the inset

$$s = \left( \frac{\mu_0 N I R^2}{2} \right)^{-2/3} .$$

The experimental value, $\mu_0 = 12.5(4) \times 10^{-7}$ Tm/A, displays great concordance with the accepted valued.

To sum up, we presented a simple and precise way to analyze the spatial dependence of magnetic fields using only a smartphone provided with acceleration and magnetic sensors. This approach can be easily extended to other experimental setups.

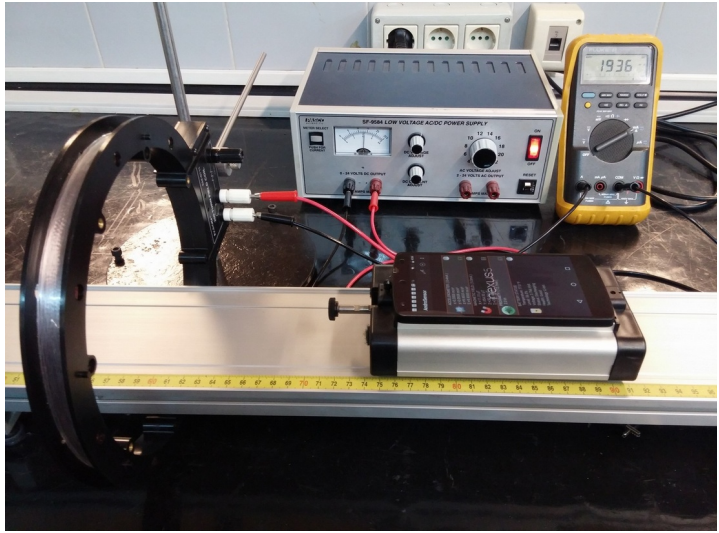

**Figure 1.** Experimental setup

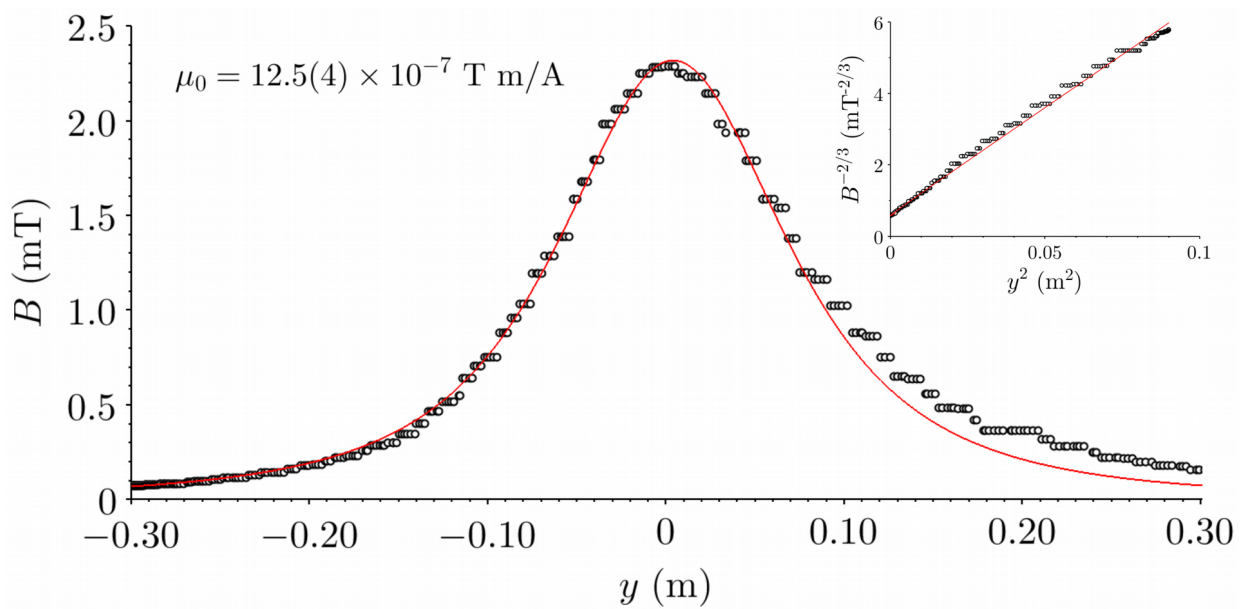

**Figure 2.** Magnetic field as a function of the axial distance. The symbols correspond to experimental results while the solid line represents the Biot-Savart law. The linear fit displayed in the inset was used to obtain $\mu_0$.